\documentclass[USenglish]{article}

\usepackage[utf8]{inputenc}%(only for the pdftex engine)
\usepackage[small]{dgruyter}
\usepackage{microtype}

\begin{document}

  \articletype{Research Article{\hfill}Open Access}

  \author*[1]{Andjelka B. Kova{\v c}evi{\'c}}
 
\author[2]{Luka {\v C}. Popovi{\'c}}

\author[3]{Dragana Ili{\'c}}

 \runningauthor{A. B. Kova{\v c}evi{\'c}, L. {\v C}. Popovi{\'c}, D. Ili{\'c}  }

  \affil[1]{Department of astronomy, Faculty of Mathematics, University of Belgrade, Studentski trg 16,11000 Belgrade, Serbia, E-mail: andjelka@matf.bg.ac.rs}

  \affil[2]{Astronomical observatory, Volgina 7, 11060 Belgrade, Serbia, E-mail: lpopovic@aob.rs}

\affil[3]{Department of astronomy, Faculty of Mathematics, University of Belgrade, Studentski trg 16,11000 Belgrade, Serbia, E-mail: dilic@matf.bg.ac.rs}

  \title{\huge Two-dimensional correlation analysis of periodicity in active galactic nuclei time series}

  \runningtitle{Analysis of periodicity in active galactic nuclei time series}

  %\subtitle{...}

  \begin{abstract}
{ The active galactic nuclei (AGN) are among the most
powerful sources with an inherent, pronounced and random variation of brightness.
The randomness of their time series  is so subtle as to blur the border between aperiodic fluctuations and noisy oscillations. This poses challenges to analysing of such time series because neither visual inspection nor pre-exisitng methods can identify well oscillatory signals in them. Thus, there is a need for  an objective method for periodicity detection. Here we review our a new data analysis method that combines a two-dimensional correlation (2D)  of time series with the powerful methods of  Gaussian processes. To demonstrate the utility of this technique, we apply it to two example problems which were not exploited enough: damped rednoised artificial time series mimicking AGN time series and newly published observed time series of changing look  AGN (CL AGN) NGC 3516. The method successfully detected periodicities in both types of time series.  Identified periodicity  of $\sim 4$ yr   in NGC 3516    allows us to speculate that if the thermal instability formed in its accretion disc (AD) on a time scale resembling detected periodicity then  AD radius could be  $\sim 0.0024$ pc. }
\end{abstract}
%  \classification[PACS]{}
 % \communicated{...}
 % \dedication{...}

  \journalname{Open Astronomy}
\DOI{DOI}
  \startpage{1}
  \received{..}
  \revised{..}
  \accepted{..}

  \journalyear{2019}
  \journalvolume{}
%  \journalissue{1}

\maketitle
\section{Introduction}
Active galactic nuclei (AGNs) vary on time-scales ranging from minutes and hours to years over the entire electromagnetic spectrum,  with no apparent indications of periodicities. 
Nevertheless, in recent years an increasing number of reports on AGN periodicities have been published \citep[see e.g.][]{2015MNRAS.453.1562G, 2015Natur.518...74G, 2016ApJ...833....6L, 2016MNRAS.463.2145C,2018MNRAS.475.2051K,
 doi.org/10.3847/1538-4365/ab0ec5,10.3847/1538-4357/aaf731}, suggesting that supermassive binary black holes  might be detected  through periodicity of their observed time series (i.e. light curves, \cite[see][and references therein]{10.1088/1361-6382/ab0587}).

Establishing  a method for  recurrent patterns  detection  in the  AGN time series is  an important step towards this goal. 
Many methods have been  designed for estimating this  periodicity
 \citep[for an excellent review see][]{doi.org/10.1093/mnras/stt1206}.
 These methods share a number of commonalities, as well as differences. Most of them are some variant of Fourier analysis \citep{doi.org/10.3847/1538-4365/aab766} which has restrictive assumptons: equally spaced observations, the time series is stationary,  homoscedastic Gaussian noise with purely periodic signals (i.e. sinusoidal shape). Wavelet analysis does not assume stationarity and is therefore able to detect amplitude and period changes over time.  Usually in all Fourier based methods  peaks which are  indicating periodicity can overlap.
However, the Fourier transform, the wavelet transform and related period estimation techniques can not tell about the presence of coordinated or independent changes among signals, as well as about relative directions of signal intensity variations.
Our hybrid method based on two-dimensional  (2D)  correlation analysis were devised to deal with above issues \citep{2018MNRAS.475.2051K}.

We aim to further  illustrate the performance and application of this 2D hybrid method  on synthetic data, where results can be judged carefully and also on  observed data, where new insights can be gained. 
We present computations of 2D correlation maps of  damped rednoised artificial time series  and newly published long-term monitored time series  of a changing look (CL) AGN NGC 3516 \citep{2019MNRAS.485.4790S}. 
There is some indication that better sampled AGN light curves can be modeled as  damped harmonic oscillator perturbed by  coloured noise \citep{doi.org/10.1093/mnras/stx1420}. 
CL AGNs are objects showing the  dramatic variability of the emission line profiles and the change of classification type within very short time interval (from days to years). 
 Periodical variability has been discussed for some  well-known CL AGN such as NGC 4151 
\citep[see][]{doi.org/10.1088/0004-637X/759/2/118, 2018MNRAS.475.2051K}, NGC 5548 \citep[see][]{2016ApJS..225...29B, 2018MNRAS.475.2051K} within the context of supermassive binary black hole candidate and  pointed out as possibility for  typical CL AGN NGC 2617 \citep[see][]{Okn18}. It would be interesting to know if CL AGN  variability is periodic, since it can be a consequence of tidal disruption events \citep{doi.org/10.1093/mnras/stw2130} or recoiling supermassive black hole \citep{doi.org/10.3847/1538-4357/aac77d}.

\section{Data and Method}

To demonstrate the utility of this technique, we apply it to two example problems which were not exploited enough.

\begin{figure}
\includegraphics[width=0.9\linewidth]{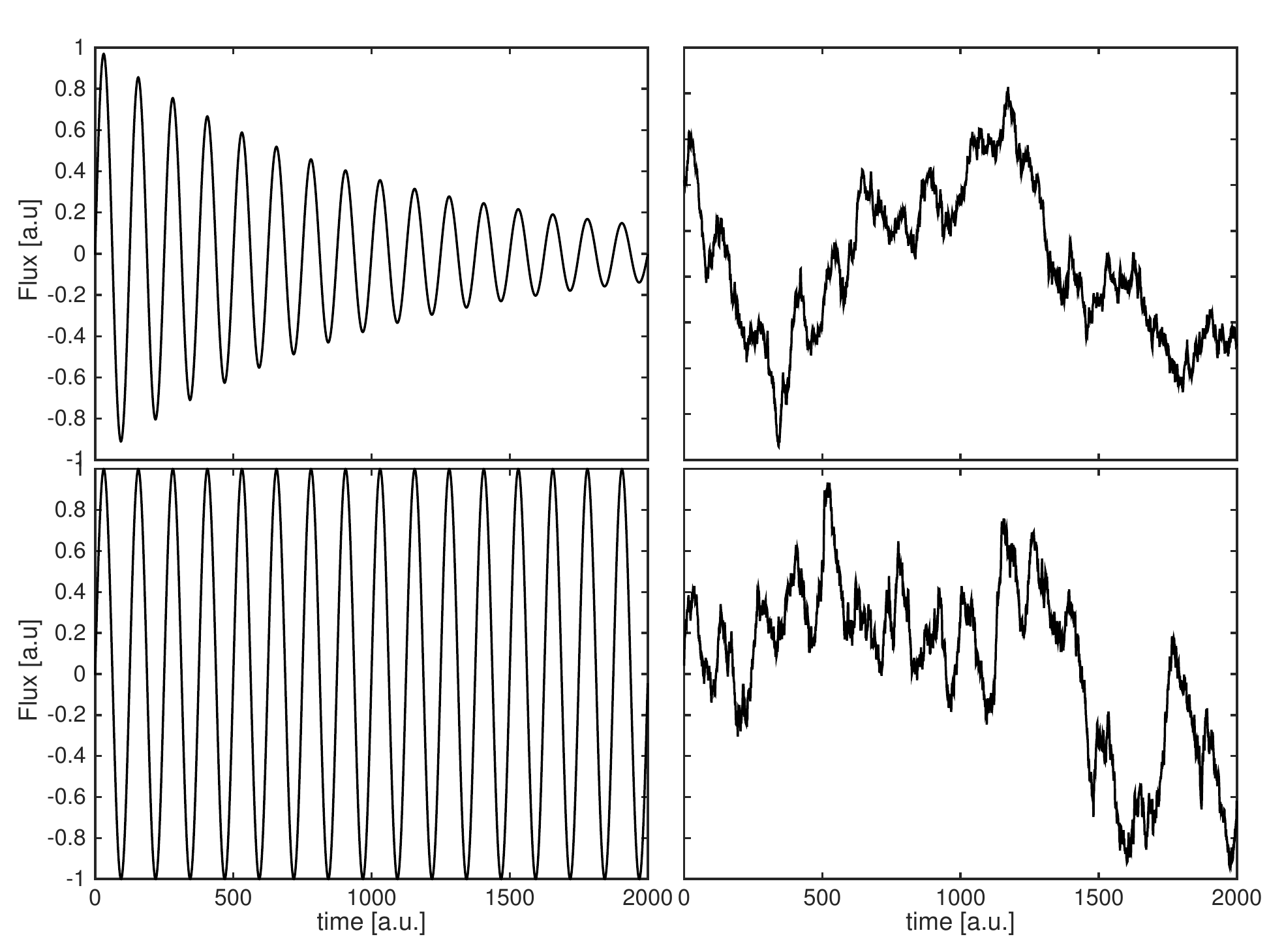}
\caption{Synthetic  light curves. A  damped sinusiod (upper left) and its rednoise corrupted form (upper right);   the same as above but for pure sinusoid (bottom left and right). Time and flux are given in arbitrary units. \label{fig1}}
\end{figure}

Interestingly,\cite{doi.org/10.1093/mnras/stx1420} made estimate that  availability of better sampled  data in the future will necessitate more sophisticated models for AGN light curves as it is damped harmonic oscillator perturbed by  colored noise.
Thus, we synthetised artificial damped sinusoid signal corrupted by red noise.
Figure \ref{fig1} (left panels)  shows both damped sinusoid and normal sinusoid  of period 125 arbitrary units [a.u.].
Normal sinusoid is symmetric with respect to time axis. But by introducing damped motion we brake this symmetry. Although the symmetry can be broken in ways that make it difficult to recognize  or reconstruct periodicity,  it is there nonetheless. And, of course, the more dramatic and complex the nature and magnitude of the damper, the more complex the task of identifying the original periodicity (symmetry). We perturbed  both signals with red noise (see right panels) so that the original signal patterns are no longer able to be recognized. Precise mathematical description of red noise  corruption of a signal is given in  \cite{2018MNRAS.475.2051K}

In a recent study of CL AGN NGC 3516 \citep{2019MNRAS.485.4790S}, authors applied Lomb Scargle periodogram in order to detect periodicity in observed light curves of this object. However,  potential  periodic signals  were of small significance.  Thus, we applied our method  to the long-term monitored  continuum, H$\alpha$ and H$\beta$ fluxes covering 22 yr (from 1996 to 2018). The data are presented and    carefully described in  \cite{2019MNRAS.485.4790S} (see their Figure 5). Thus,  we will not repeat the details  here.

\begin{figure}
\includegraphics[width=0.9\linewidth]{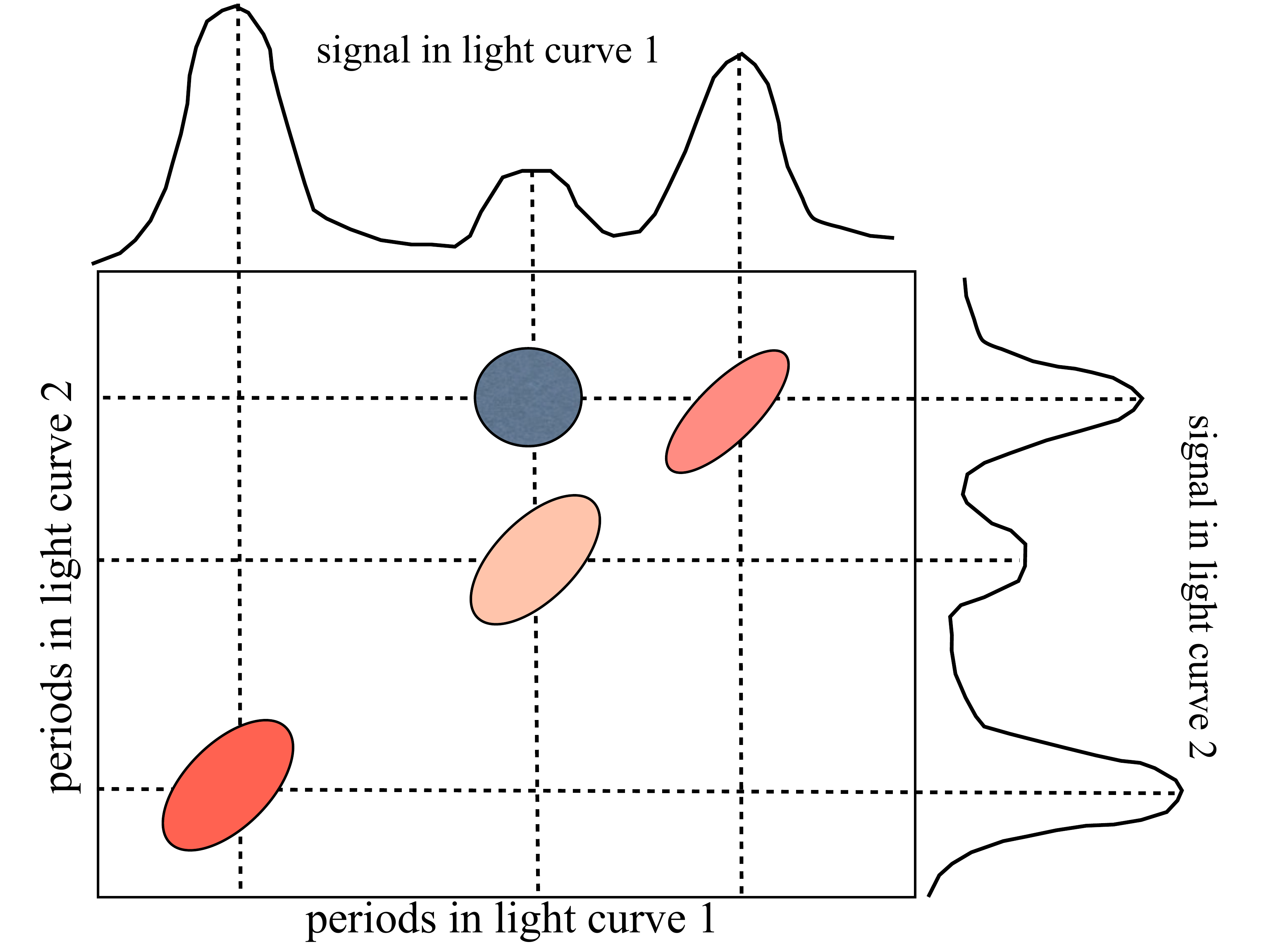}
\caption{A schematic example of 2D correlation map. Signals present in light curve 1 and 2 are given on the top and right of the map. Cross peaks off diagonal are not detected in AGN light curves.\label{fig2}}
\end{figure}

Here we recapitulate key aspects of our hybrid method,  which is discussed in details in \cite{2018MNRAS.475.2051K}.
A conversion of set of light curves  to  2D correlation maps is relatively easy, providing rich information about the presence of coordinated or independent signals as well as relative directions of signal variations. 
Some notable features of our 2D hybrid method are:  simplification of complex spectra consisting of many overlapped peaks, enhancement of apparent spectral resolution by spreading peaks over the second dimension, and establishment of direction of changes in signal through correlation coefficients.
Some generic properties of  correlation map are marked on Figure \ref{fig2}.
Our hybrid method produces a contour map of correlation intensity on a period plane defined by two independent period axes corresponding to the two light curves. Peaks on the main diagonal represent the simultaneous change in signals at the same period. Cross peaks located at the off diagonal positions  represent the simultaneous change in signals at two different periods. However, we have never observed these cross peaks in objects  we analyzed up to now. Positive correlation value means  the two periodic signals increase or decrease together in the same direction.

\subsection{Results and Discussion}

\begin{figure}
\includegraphics[width=0.9\linewidth]{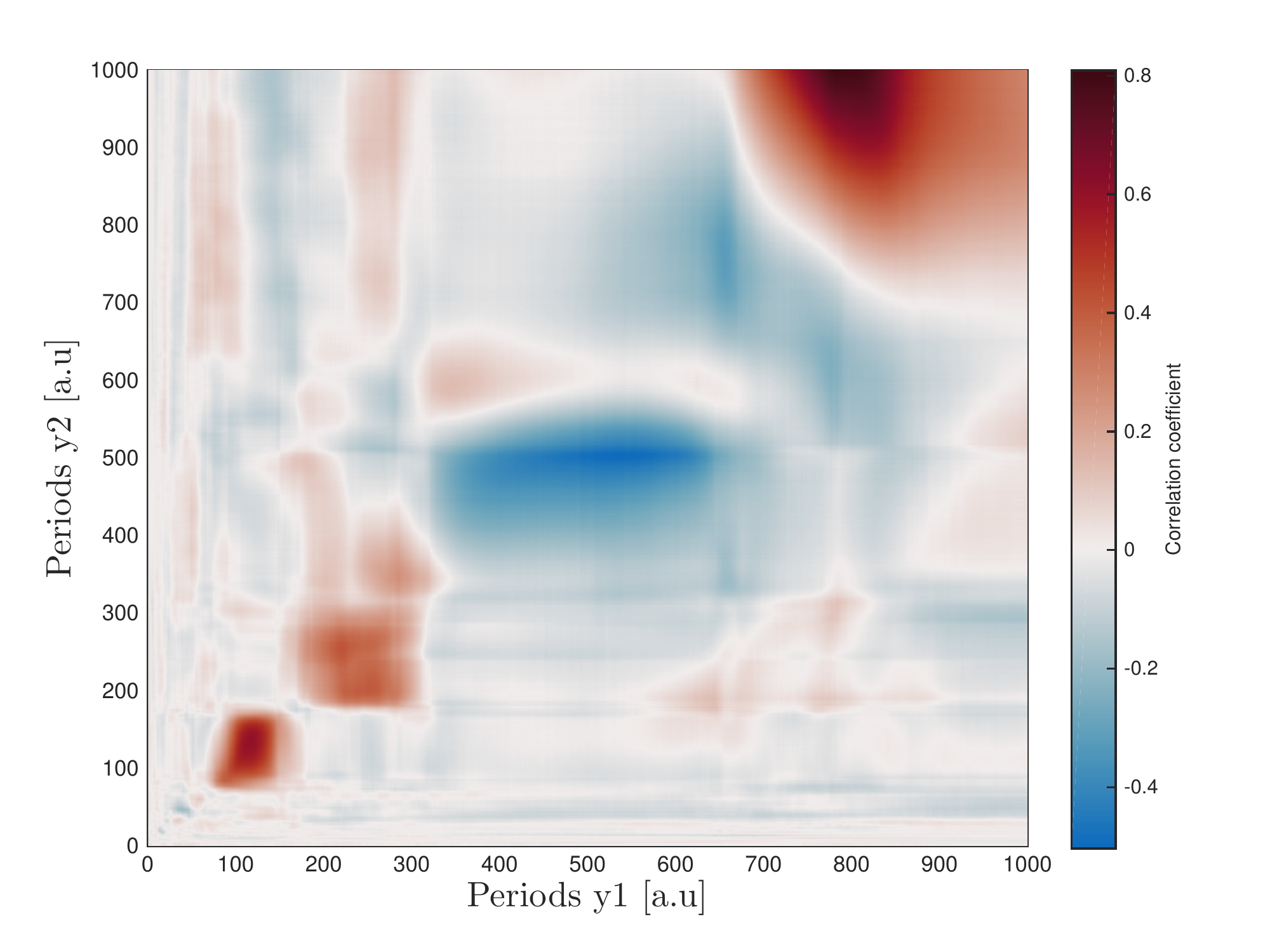}
\caption{A 2D correlation map of  sinusoidal (y1) and damped sinusoidal (y2) signals, with matching period of 125 [a.u]  and corrupted with random noise. The peak at lower left of the diagonal  indicates period of $122\pm 26$ [a.u]. Due to change in width of waves in damped sinusoid, periodic signals  are not in perfect phase or antiphase which is indicated by  presence of Nyquist period  in the upper right corner and two weakly correlated islands in the middle ($\sim 250$[a.u] and $\sim 500$ [a.u]).\label{fig3}}
\end{figure}

Firstly, we present a 2D correlation map (see Figure \ref{fig3}) of  a pair of synthetic curves (a sinusoid and damped sinusoid  of period 125 [a.u] corrupted  with red noise, see right column of Figure \ref{fig1})  to illustrate this approach for establishing the presence of common periodicities.

Because of no overlap in this map, the peaks can be readily assigned to specific periodic signals.
Peak appearing at the lower left of diagonal positions of the 2D map indicate a clear positive correlation at the period of $122\pm 26$[a.u.] with  correlation coefficient  $>0.8$.  However, the peak in upper right corner is related to the Nyqist frequency and is not relevant. Clearly the width of waves in damped sinusoid is changing over time and they become wider then waves of sinusoid.  As a consequence the signals are not in perfect phase. Due to this,   very weakly correlated positive ($\sim 250$ [a.u]) and negative  ($\sim 500$ [a.u])   islands   appear between lower left and upper right correlation islands.

\begin{figure}
\includegraphics[width=0.9\textwidth]{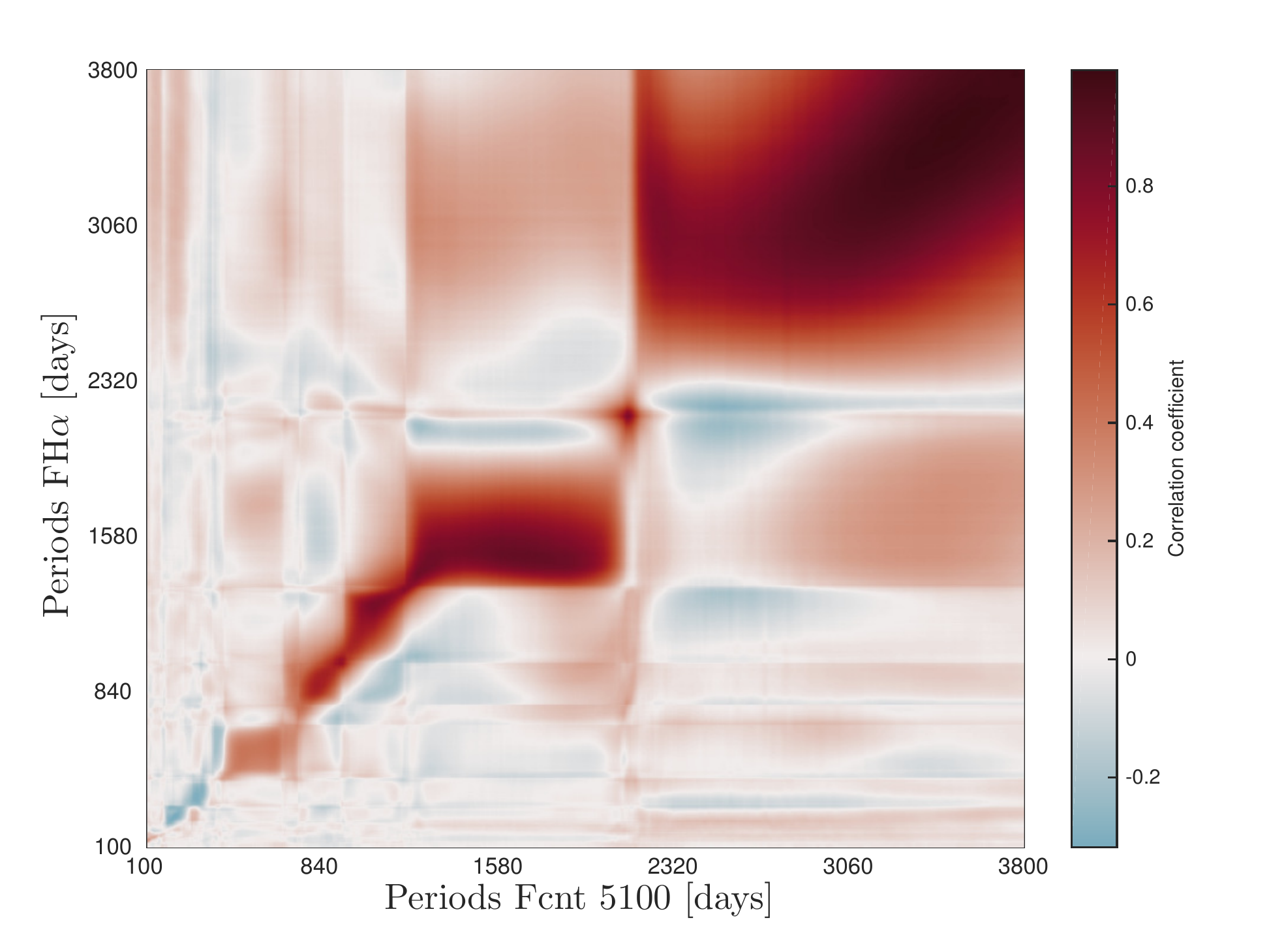}
\caption{A 2D correlation map of  continuum and H$\alpha$ emission line of NGC 3516, with matching period of  $1580\pm 743$ days.  Upper right peak corresponds to whole period covered by observations.\label{fig4}}
\end{figure}

\begin{figure}
\includegraphics[width=0.9\textwidth]{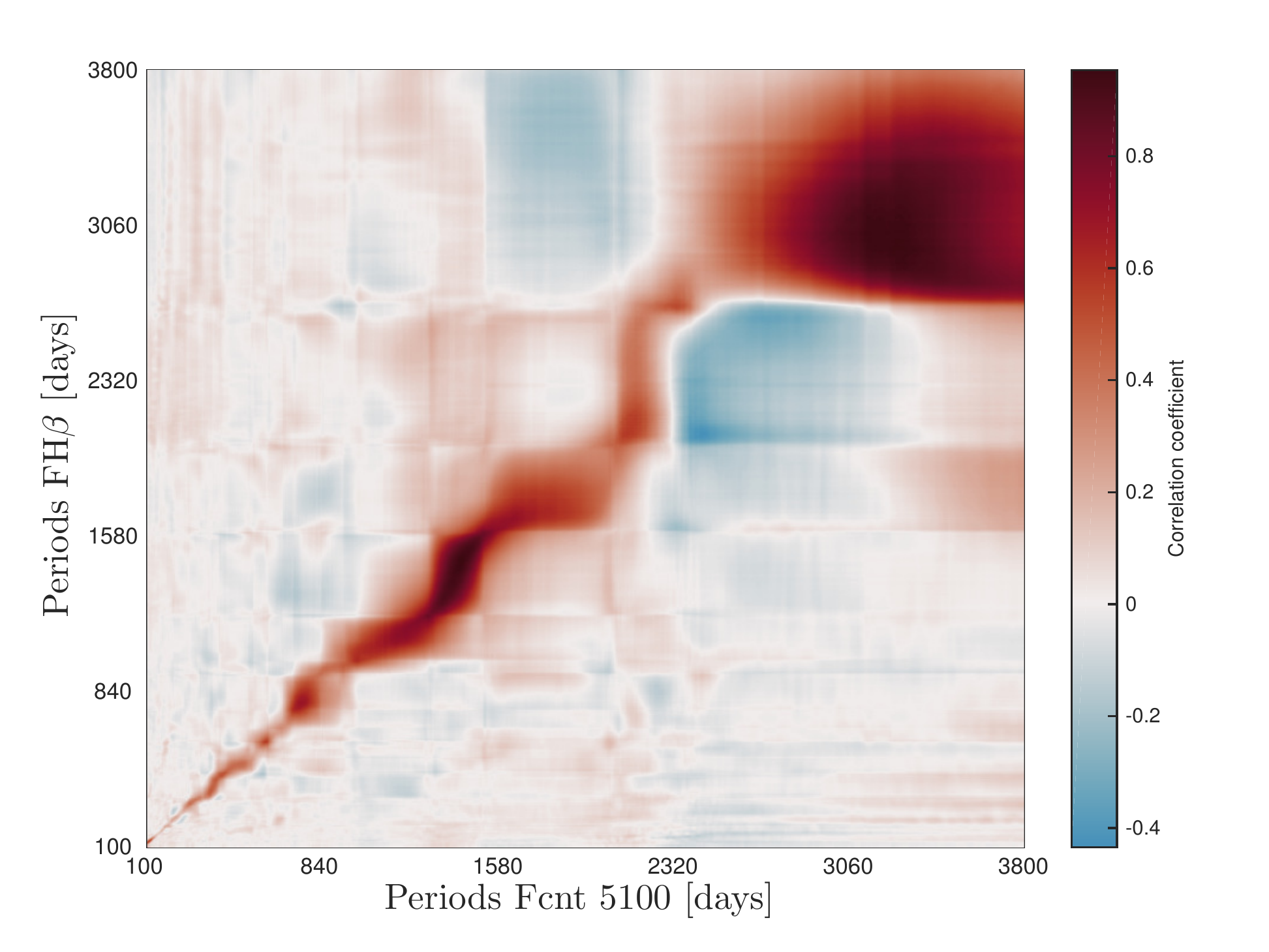}
\caption{The same as Figure \ref{fig4} but for continuum and H$\beta$ emission line of NGC 3516. Detected period is $1385\pm 128$ days.\label{fig5}}
\end{figure}

Furthermore, we also applied our hybrid method to the long-term monitored continuum, H$\alpha$ and H$\beta$ light curves  of CL AGN NGC 3516.
Our long-term observations are covering 22 years, however
due to lack of  data after year 2007, here we used only 10-year long part of the light curve up to 
2007 (MJD 54500).
 Resulting 2D correlation maps are shown in 
Figures \ref{fig4} and \ref{fig5}, allowing effortless identification of links between periodicities in continuum and H$\alpha$, as well as continuum and H$\beta$, respectively.

Unexpectedly, the strong relationships are present at  periods $1580\pm743$  days   (i.e. $4.32\pm 2.04$yr  \cite[compare to Figure 11 in][]{2019MNRAS.485.4790S}) and $1385\pm 128$ days (i.e. $3.8\pm 0.4$yr, see  Figure \ref{fig5}).  
Calculated correlation maps provide an even clearer picture of the time-dependent changes of periodicities in the light curves than Lomb periodogram \citep[compare to Figure 11 in][]{2019MNRAS.485.4790S}.

These periodicities are similar to each other, indicating that continuum, H$\alpha$ and H$\beta$ fluctuate with similar periodicity characteristics. It seems that periodicity in continuum, triggers similar variation in H$\alpha$ and H$\beta$ emission lines. 

 Perturbed AGN accretion disc can be used as an explanation of continuum flux and corresponding spectral variability \citep[see e.g.][]{2014A&A...572A..66P}. It is believed  that such perturbations induce  the thermal instability in the disc  which time scale is
  \citep[see][]{2008ApJ...677..884L}
 \begin{equation}
 t_\mathrm{th}=5\frac{0.1}{\alpha}M_\mathrm{BH}[10^{8} M_{\odot}]\sqrt{(\frac{r_\mathrm{d}}{10^{3} r_\mathrm{g}})^3} \mathrm{yr},
 \label{instab}
 \end{equation}
where $r_\mathrm{g}=GM_\mathrm{BH}/c^2 $ is a gravitational radius and $r_\mathrm{d}$ is accretion disc dimension.
If the thermal instability had formed in the disc of NGC 3516 on a time scale resembling detected periodicity ( $\sim 4$ yr) we can infer the disc dimension using Eq.  \ref{instab}.
Substituting the standard value for $\alpha=0.1$ \citep[see][and references therein]{doi.org/10.3847/1538-4357/aac77d}, the
 black hole mass of NGC 3516 $M_\mathrm{BH}=4.73 \times 10^{7} M_{\odot}$\citep{2019MNRAS.485.4790S}  in Eq. \ref{instab} and adopting detected periodicity as a thermal scale, we obtain  accretion disc radius $\sim 0.0024$ pc. This agrees well with 
\cite{doi.org/10.1051/0004-6361:20020724} prediction of accretion disc dimension for this object
(between 0.004 pc and 0.018 pc) and with indices of dimension
variability of emitting region  \citep[see detailed discussion in][]{2019MNRAS.485.4790S}.
In order to  test hypothesis about recoiling supermassive black hole the broad emission lines velocities offsets should be calculated and modeled \citep[see][for detailed explanation]{doi.org/10.3847/1538-4357/aac77d}.
To confirm periodicity and explain variability of this object, we need more spectral and photometric observations covering at least another ten years.
 Although this study was carried out with a relatively small time coverage of NGC 3516 time series, it indicates that  NGC 3516  is interesting object for periodicity detection, and  its  monitoring should continue.

\section{Conclusion}

In the present study, we show  a new application of  our hybrid method for periodicity detection in AGN time series. We extended the results obtained in \cite{2018MNRAS.475.2051K, 10.3847/1538-4357/aaf731}, via the analysis of synthetic sinusoidal and damped sinusoidal signals corrupted with red noise  as well as observed  continuum, H$\alpha$ and H$\beta$ light curves of  CL AGN NGC 3516. Our hybrid method successfully recovered the period in synthetic time series (i.e. light curves). Further, it detected periods of $1580\pm743$ days in the continuum and H$\alpha$ as well as $1385\pm 128$ days in the the continuum and H$\beta$ of NGC 3516. If the thermal instability had formed in the accretion disc of NGC 3516 on a time scale resembling detected periodicity ($\sim 4$ yr) we inferred that accretion disc radius is $\sim 0.0024$ pc. This agrees well with \cite{doi.org/10.1051/0004-6361:20020724} prediction of accretion disc model for this object and with \cite{2019MNRAS.485.4790S}  evidence  of its dimension  variability.

Both experimental results show the robustness of our method against problems of damped oscillations corrupted with red noise and complex time series of CL AGN NGC 3516. However, in order to validate periodicity detection in   NGC 3516 time series,  a further decade long-term monitoring is needed.

Acknowledgement.
This work was 
present at 12th SCSLSA in the special session:
{\it Broad lines in AGNs: The physics of emission gas in the vicinity of 
super-massive black hole}  (In memory of the life and work of dr Alla 
Ivanovna Shapovalova). This work is supported by project (176001) Astrophysical Spectroscopy of Extragalactic Objects.


\begin{thebibliography}{99}

\bibitem[Graham et al.(2015a)]{2015MNRAS.453.1562G}
Graham, M. J., Djorgovski, S. G., Stern, D., et al. 2015a, MNRAS, 453, 1562--1576

\bibitem[Graham et al.(2015b)]{2015Natur.518...74G}
Graham, M. J., Djorgovski, S. G., Stern, D., et al. 2015b, Nature, 518, 74--76

\bibitem[Liu et al.(2016)]{2016ApJ...833....6L}
Liu, T., Gezari, S., Burgett, W., et al. 2016, ApJ, 833, 6, 13pp

\bibitem[Charisi et al.(2016)]{2016MNRAS.463.2145C}
Charisi, M., Bartos, I., Haiman, Z., et al. 2016, MNRAS, 463, 2145--2171

\bibitem[Kova{\v c}evi{\' c} et al.(2018)]{2018MNRAS.475.2051K}
Kova{\v c}evi{\' c}, A. B., P{\'e}rez-Hern{\'a}ndez, E., Popovi{\' c}, L. {\v C}., Shapovalova, A. I.,  Kollatschny, W.,  Ili{\' c}, D.  2018, MNRAS, 475, 2051--2066

\bibitem[Li et al.(2019)]{doi.org/10.3847/1538-4365/ab0ec5}
Li, Y. R.,  Wang, J.-M.,  Zhang, Z.-X, Wang, K. et al. 2019, ApJSS, 241, 14pp

\bibitem[Kova{\v c}evi{\' c} et al.(2019)]{10.3847/1538-4357/aaf731}
Kova{\v c}evi{\' c}, A. B., Popovi{\' c}, L. {\v C}., Simi{\'c}, S.,  Ili{\' c}, D., 2019, ApJ, 871, id. 32, 1--11

\bibitem[Barack et al.(2019)]{10.1088/1361-6382/ab0587}
Barack, L., Cardoso, V., Nissanke, S., Sotiriou, T. P., Askar, A., Belczynski et al. 2019, Classical and Quantum Gravity, 36, 1--272

\bibitem[Graham et al.(2013)]{doi.org/10.1093/mnras/stt1206}
Graham, M. J., Drake, A. J.,  Djorgovski, S. G.,  Mahabal, A. A., Donalek, C. 2013, MNRAS, 434, 2629--2635

\bibitem[VanderPlas (2018)]{doi.org/10.3847/1538-4365/aab766}
 VanderPlas, J. T. 2018, ApJS, 236, 22pp


\bibitem[Shapovalova et al.(2019)]{2019MNRAS.485.4790S}
Shapovalova, A. I., Popovi{\' c}, L. {\v C}., Afanasiev, V. L., Ili{\' c}, D., Kova{\v c}evi{\' c}, A. B., Burenkov, A. N., Chavushyan, V. H., Mar{\v c}eta-Mandi{\' c}, S. et al. 2019, MNRAS, 485, 4790--4803

\bibitem[Kasliwal et al.(2017)]{doi.org/10.1093/mnras/stx1420}
Kasliwal, V. P.,   Vogeley, M. S.,  Richards, G. T. 2017, MNRAS, 470,  3027–-3048


\bibitem[Bon et al.(2012)]{doi.org/10.1088/0004-637X/759/2/118}
Bon, E.,  Jovanovi{\'c}, P., Marziani, P., Shapovalova, A. I., Bon, N.,  Jovanovi{\' c}, V. B., Borka, D., Sulentic, J., Popovi{\' c}, L. {\v C}. 2012, ApJ, 759, 8pp

\bibitem[Bon et al.(2016)]{2016ApJS..225...29B}
Bon, E., Zucker, S., Netzer, H., Marziani, P., Bon, N., Jovanovi{\' c}, P., Shapovalova, A.I.,  Komossa, S., Gaskell, C. M., Popovi{\' c}, L. {\v C}. et al. 2016, Ap.JS, 225, 15pp

\bibitem[Oknyansky et al.(2018)]{Okn18}
Oknyansky, V. L., Malanchev, K. L., Gaskell, C. M.  In Proceeding of the POS: Revisiting Narrow-Line Seyfert 1 Galaxies and Their Place in the Universe, Padova, Italy, 9–13 April 2018., 1--5

\bibitem[Xiang-Gruess et al.(2016)]{doi.org/10.1093/mnras/stw2130}
Xiang-Gruess, M., Ivanov, P. B., Papaloizou, J. C. B. 2016, MNRAS, 463, 2242--2264

\bibitem[Kim et al.(2018)]{doi.org/10.3847/1538-4357/aac77d}
 Kim, D. C.,  Yoon, I.,   Evans, A. S. 2018, ApJ, 861, 1--10

\bibitem[Popovi{\' c} et al.(2014)]{2014A&A...572A..66P}
Popovi{\' c}, L. {\v C}., Shapovalova, A. I., Ili{\' c}, D., Burenkov, A. N., Chavushyan, V. H., Kollatschny, W., Kova{\v c}evi{\' c}, A. et al. 2014, A\&A,  572, id.A66, 17pp

\bibitem[Liu et al.(2008)]{2008ApJ...677..884L}
Liu, H. T., Bai, J. M., Zhao, X. H., Ma, L. 2008, ApJ, 677, 884--894

\bibitem[Popovi{\' c} et al.(2002)]{doi.org/10.1051/0004-6361:20020724}
Popovi{\' c}, L. {\v C}., Mediavilla, E. G.,  Kubi{\v c}ela, A.,  Jovanovi{\'c}, P. 2002, A\&A, 390, 473--480



\end{thebibliography}
\end{document}